\documentclass{elsart}

\usepackage{graphicx}
\usepackage{amssymb}

\begin{document}

\begin{frontmatter}



\title{Magnetism and superconductivity of heavy fermion matter}


\author[CEA]{J. Flouquet},
\author[CEA]{G. Knebel},
\author[CEA]{D. Braithwaite},
\author[IMR]{D. Aoki},
\author[CRTBT]{J.P. Brison},
\author[CEA]{F. Hardy},
\author[CEA]{A. Huxley},
\author[CEA]{S.Raymond},
\author[CEA]{B. Salce},
\author[LCMI]{I. Sheikin},

\address[CEA]{D\'{e}partement de la Recherche Fondamentale sur la Mati\`{e}re Condens\'{e}e, SPSMS, CEA Grenoble, 17 rue des Martyrs, 38054 Grenoble Cedex 9, France}
\address[IMR]{IMR, Tohoku University,
Oarai, Ibaraki 311-1313, Japan}
\address[CRTBT]{Centre de Recherche sur les Tr\`{e}s Basses Temp\'{e}ratures, CNRS, 38042 Grenoble Cedex 9, France}
\address[LCMI]{Grenoble High Magnetic Field Laboratory, MPI-FKF/CNRS, BP166, 38042 Grenoble Cedex 9, France}

\begin{abstract}
The interplay of magnetism and unconventional superconductivity (d singlet wave or p triplet wave) in strongly correlated electronic system (SCES) is discussed with recent examples found in heavy fermion compounds. A short presentation is given on the formation of the heavy quasiparticle with the two sources of a local and intersite enhancement for the effective mass. Two cases of the coexistence or repulsion of antiferromagnetism and superconductivity are given with CeIn$_3$ and CeCoIn$_5$. A spectacular example is the emergence of superconductivity in relatively strong itinerant ferromagnets UGe$_2$ and URhGe. The impact of heavy fermion matter among other SCES as organic conductor or high $T_C$ oxide is briefly pointed out.
% Text of abstract
\end{abstract}

\begin{keyword}
heavy fermion, superconductivity, antiferromagnetism, ferromagnetism
\PACS 71.27.+a; 74.70.Tx; 74.20.Mn
\end{keyword}
\end{frontmatter}

% main text
%\section{Conventional case}
%\label{}

%\begin{document}

%\title
%Magnetism and superconductivity of heavy fermion matter.

%\authors
%J. Flouquet, G. Knebel, D. braithwaite, D. Aoki, J.P. Brison, F. Hardy, A. Huxley, I. Sheikin, S. Raymond, %B. Salce

\section{Conventional case}
The interplay of magnetism and superconductivity  covers quite different situations. In the past decade, the novelty comes with the discovery of new materials where the electronic correlations are  strong: organic conductor \cite{Jerome1980}, high $T_C$ oxides \cite{Bednorz1986}, or heavy fermion systems \cite{Steglich1979}. The localisation or the motion of the quasiparticles leads often to competing ground states with  transitions from insulator to metal or from well ordered magnetic phases to a paramagnetic (PM) state associated or not with the appearance of superconductivity (S) \cite{Flouquet2005}. In these strongly correlated electronic systems (SCES) quite different facets can be stressed : for example for superconductivity the pairing mechanism, the stability of superconductivity, the nature of the order parameter,  the consequences on the low energy excitations, and the feedback on the vortex matter \cite{Thalmeier2004}. Here we will focus mainly on the experimental determination of the temperature ($T$), pressure ($P$) and magnetic field ($H$) phase diagram of heavy fermion compounds (HFC). A new type of matter may occur at the boundary between two ground states which can be tuned under $P$ or $H$.

In the conventional case with weak correlations between the itinerant quasiparticles, two different baths exist: the magnetism is carried by localised moments which interact weakly with the other electronic bath formed by the Fermi sea \cite{Fisher1975,Fisher1978}. The peaceful coexistence of antiferromagnetism (AF) and superconductivity were discovered in 1975 at the University of Geneva in ternary compounds of rare earth (RE) elements and molybdenum sulfide (REMo$_6$S$_8$) \cite{Fisher1975} later in a serie of rhodium boride alloys (RERh$_4$B$_4$) \cite{Vandenberg1977,Fertig1977} and more recently in RE borocarbide \cite{Canfield1998}. Often most of these compounds are superconducting below a critical temperature $T_C \approx 2$ K to 10 K and undergo phase transition to an antiferromagnetically ordered state at $T_N \leq T_C$. The simple argument is that at the scale of the superconducting coherence length $\xi_0$ the Cooper pair will feel an average zero magnetic field so far $\xi_0$ will be larger than the magnetic period $d$. This length $d$ extends usually only to few atomic distances with also short magnetic correlation $\xi_m \leq \xi_0$. The superconducting pairing leads to the so called s-wave pairing: the spin up and spin down paired electrons have zero orbital angular momentum and the mechanism of attraction is due to the electron - phonon interaction.

A magnetic field can destroy singlet superconductivity in two ways. The orbital effect is simply the manifestation of the Lorentz force; the corresponding orbital limit $H_{C2}(0) $ at $T \rightarrow$ 0 K varies as $(m^{* 2} T_C)$, where $m^*$ is the effective mass of the quasiparticle. The paramagnetic limitation ($H_p$) occurs when a strong magnetic field attempts to align the spins of both the electrons: $H_p (0) = 1.8 T_C$  in T. In conventional superconductors, as $H_{C2}(0) \leq H_p (0)$ due to the weakness of $m^*$, the Pauli limit is often irrelevant. However in HFC with $m^* \approx 100 m_0$ ($m_0$ the free electron mass), a competition will occur between the two mechanisms at least for singlet pairing. However, for triplet superconductivity with equal spin pairing between up up or down down spins, there will be no Pauli limitation and  superconductivity can only be destroyed by the orbital effect.

In conventional superconductors as observed in ErRh$_4$B$_4$ and HoMo$_6$S$_8$, the superconductivity is destroyed by the onset of a first order ferromagnetic transition. The energy gained by the atoms carrying the ferromagnetic moment below $T_{{\rm Curie}}$ exceeds  the energy gained by the electrons as they form Cooper pairs at $T_C$ . Superconductivity cannot prevent the magnetic transition but can only modify it slightly in a narrow temperature range. Furthermore, it is very unlikely for singlet superconductivity to also survive in the ferromagnetic (FM) state because the exchange interaction $H_{ex}$ often forbids the formation of Cooper pairs  \cite{Ginzburg1957}.

\section{The heavy fermion matter}

\begin{figure}[h]
	\centering
	\includegraphics[scale=0.5,clip]{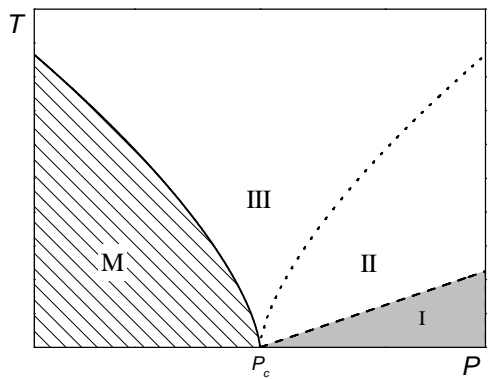}
	\caption{Magnetic phase diagram predicted for an itinerant spin fluctuation system. In the domain I, Fermi liquid properties will be achieved. In the domain II and III, non Fermi-liquid behaviour will be found. Depending on the nature of the interactions (AF or F), the contour lines I, II, III change. Here the contour is drawn for antiferromagnetism. In case of ferromagnetism, the $T_I$ line starts as $(P - P_C)^{3/2}$.}
	\label{figure1}
\end{figure}

The heavy fermion compounds are inter-metallic compounds of 4f or 5f electrons where on cooling, very large effective masses often 100 m$_0$ appears due to the weak delocalisation of the 4f or 5f particle from its site produced by the hybridisation with the other initial light itinerant electrons (s, p, d) \cite{Flouquet2005}. Basically on cooling below 10 K the remaining large magnetic entropy ($S = R \log 2$ for a doublet crystal field ground state) is transferred for the formation of heavy quasiparticle with  an effective low Fermi temperature (10 K instead of 10.000 K for noble metals). However the duality between the localised and itinerant character leads to the competition between long range magnetic order (AF or FM) and PM. The popular picture is that the switch  at $T \rightarrow 0$ K will appear at a magnetic quantum critical point (QCP) via a smooth second order phase transition. For example under pressure, that will occur at $P_C$ (figure \ref{figure1}).
 
We will not enter in the present debate on HFC descriptions.  Qualitatively, at least on the PM site, their properties are rather well described by an itinerant spin fluctuation approach with vanishing crossover temperatures $T_I$ and T$_{II}$ where respectively Fermi liquid properties are well obeyed below $T_I$ and the usual high temperature behaviour of Kondo impurities is recovered above T$_{II}$. In the temperature window T$_{III}$ - $T_I$ the so called non Fermi liquid properties are observed. The underlining question is the location of $P_C$ by respect to the pressure $P_{KL}$ where the 4f electrons participate ($P \geq  P_{KL}$) or not ($P \leq P_{KL}$) to the Fermi sea and to the pressure $P_V$ where for $P \geq P_V$ the 4f electron looses their orbital sensitivity to the local environment (the crystal field effect) due to the hybridisation with the other electrons. At least $P_C$ must be between $P_V$ and $P_{KL}$. The physical image of HFC is that, through strong local fluctuations reminiscent of the Kondo effect of a single impurity characterised by its Kondo temperature ($T_K$), the renormalised band corresponds already to a heavy particle with a band mass near $m_b \cong 50 m_0$. Antiferromagnetic fluctuations produced by the intersite coupling give only an extra factor of 2. A strong support for this statement is that a self-consistent fit of the superconducting upper critical field $H_{C2} (0)$ including orbital, Pauli plus strong coupling limits with $\lambda$ coefficient $\lambda =  \frac{m^*}{m_b}-1$ indicates a moderate value of $\lambda \approx 1.2$ ie a ratio of $m^* = 2.2 m_b$.

\begin{figure}[h]
	\centering
 \includegraphics[scale=0.5,clip]{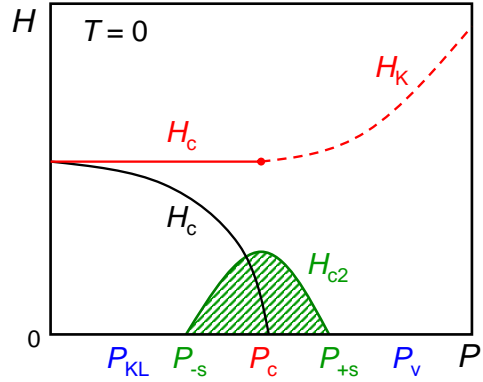}
	\caption{At $T = 0$K, characteristic pressure of HFC : $P_{KL}$ when the 4f electrons are itinerant, $P_{-S}$ and $P_{+S}$ for the onset and disappearance of superconductivity, $P_C$ the magnetic quantum critical point and $P_V$ when strong valence or orbital fluctuations exist as the Kondo temperature overpasses the crystal field splitting. $H_C$ represents the critical field from AF to PM state. Depending of the magnetic anisotropy, it ends a critical point or collapses at $P_C$. $H_K$ represents the entrance in a Kondo polarised PM state.}
	\label{figure2}
\end {figure}

For different cerium HFC - AF, a superconducting dome appears tight to $P_C$ with the appearance and disappearance of superconductivity at $P_{-S}$, $P_{+S}$ (figure \ref{figure2}). The nearly coincidence of the maxima $T^{max}_{C}$ of $T_C$ near $P_C$ is an indirect strong support for a magnetic origin of the superconducting pairing. At P$_C$, slow magnetic fluctuation occurs : $T_I$ vanished linearly with $P - P_C$ for AF ; $m^*/m_b$ reaches its maxima. As the characteristic temperatures ($T_N$, $T_I$) collapses at $P_C$, their strong pressure dependence leads to a huge corresponding Gr\"{u}neisen coefficient 
$\Omega^* =- \frac{\partial \log T_I} {\partial \log T_V}$  ($V$: volume molar). Through the Maxwell relation, a huge electronic thermal expansion  occurs basically $\alpha \approx m^{*2}$ for $P \geq P_C$ \cite{Flouquet2005}. The combined effects of large spin and density fluctuations can induce a Cooper pairing with an unconventional order parameter \cite{Thalmeier2004} (at least different from the previous s wave case) since the local strong coulomb repulsion precludes the simultaneous presence of the Cooper pair on a given site: singlet d wave or triplet pairing have been reported.

\begin{figure}[h]
	\centering
	\includegraphics[scale=0.35,clip]{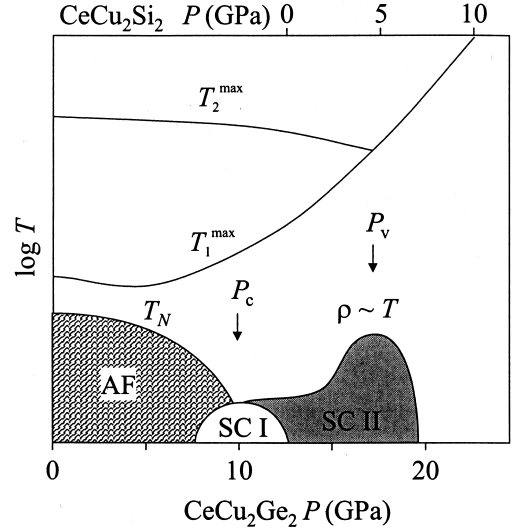}
	\caption{Schematic $P-T$ phase diagram of CeCu$_2$(Si/Ge)$_2$ showing the two critical pressures $P_C$  and $P_V$. Near $P_C$ , where the antiferromagnetic ordering temperature $T_N \rightarrow 0$, superconductivity in region SC I is governed by antiferromagnetic spin fluctuations.  Around $P_V$, in the region SC II, valence fluctuations provide the pairing mechanism and the resistivity is linear in temperature. The temperatures $T^{max}_{1}$ , and $T^{max}_{2}$ of the temperature of the maxima of the resistivity merge at a pressure coinciding with $P_V$. (\cite{Holmes2004}).}
	\label{figure3}
\end {figure}

There is now an extensive literature on superconductivity induced by spin fluctuation \cite{Moriya2003,Chubukov2003}. The energy window between the frequencies $\hbar\omega_{sf} = k_B T_I$ and $\hbar\omega_K = k_B T_K$ play a key role in the pairing. Even if the transition at $P_C$ is not of second order but weakly first order there will be no strong effect on the appearence of superconductivity. When $P_C \approx P_V$ (CeIn$_3$, CePd$_2$Si$_2$, CeRh$_2$Si$_2$ cases) $T^{max}_{C}$ seems located near $P_C$. For CeCu$_2$Si$_2$ or CeCu$_2$Ge$_2$ where $P_C \leq P_V$ two different regimes occur in the pressure variation of $T_C$ (figure \ref{figure3}), \cite{Holmes2004,Yuan2003}. This suggests two different mechanisms for the Copper pairing mediated by spin or valence fluctuations.

The magnetic field leads to the creation of a vortex lattice with the mixed phase of normal and superconducting components. In the special situation of SCES, it may lead to new phenomena : (i) the restoration of specific magnetic properties inside the vortex core as recently reported in high $T_C$ superconductors \cite{Luke2001}, (ii) a displacement of the frontier between AF and S phases when it occurs.

In this very narrow band an extra effect may occur due to the field variation of the ground state itself but also in the nature of the magnetic fluctuation. In the ($H$, $P$) phase diagram at 0 K (figure \ref{figure2}), the superconducting upper critical field $H_{C2} (0)$  must be located by comparison to characteristic fields for the magnetism : $H_C$ for the transition from AF to PM states below $P_C$, $H_m$ inside the PM phase for the entrance in the polarised PM state (PPM). Above $H_m$, the static ferromagnetic component increases continuously but the $q$ dependence of the dynamical response is flat. If the local magnetism is of Ising type (the magnetic anisotropy axial) $H_m$ is the continuation above $P_C$ of $H_C$ which ends up at a critical point at $P_C$ : $H_m$ is roughly equal to the Kondo field $H_K = k_B T_K/9 \mu_B$ \cite{Flouquet2005}. For a planar or Heisenberg local spin, $H_C$ will collapse at $P_C$ and thus there will be a large field range above $P_C$ from $H = 0$ to $H_K$ where AF and FM fluctuations complete. By contrast in the Ising case, both interactions will interfer in a narrow field window around $H_m \approx H_K$ \cite{Flouquet2002}.

\section{The superconducting magnetic boundaries}

\begin{figure}[b]
\centering
\includegraphics[scale=0.6,clip]{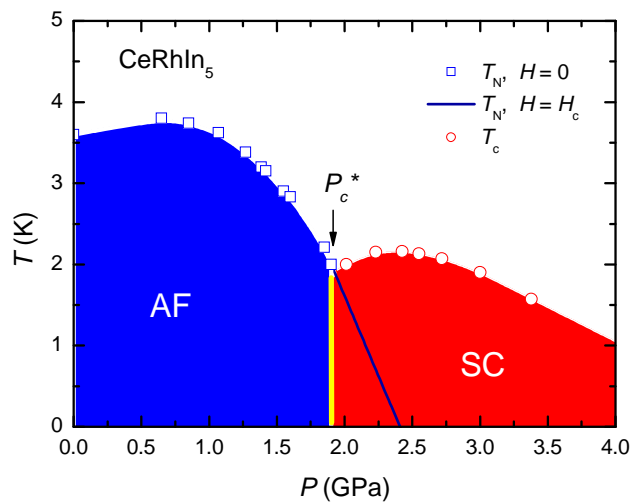}
\caption{Schematic $(T, P)$ phase diagram of CeRhIn$_5$. The points observed by resistivity, susceptibility and specific heat below $P^{*}_{C}$ characteristic of an extra inhomogeneous media have been removed (see figure 9). For $H \geq H_{C2} (0)$  and $P^{*}_{C} \leq P \leq P_C$, AF will continuously reappear up to reach $T_N \rightarrow 0$ K near $P_C$ (see \cite{Blount1990,Kivelson1998}).}
\label{figure4}
\end{figure}

\begin{figure}[t]
	\centering
\includegraphics[scale=0.6]{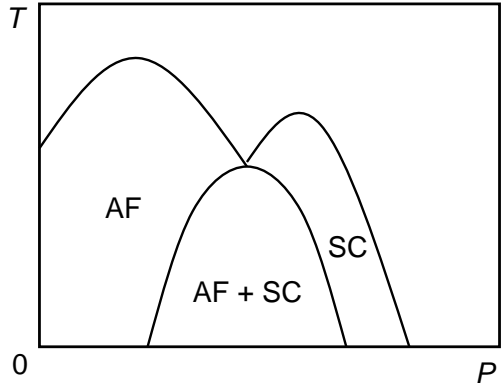}
	\caption{Schematic $(T, P)$ phase diagram of the interplay between AF and S when a coexistence domain exists as it seems to occur for different HFC. However $T^{Max}_{C}$ occurs often close to $P_C$ when $T^{Max}_{N} \approx 10 T^{Max}_{C}$.\cite{Kivelson1998,Demler2004}}
	\label{figure5}
\end{figure}

The thermodynamics of the interplay between magnetism and superconductivity was often neglected for the debate on the mechanism for the superconducting pairing and on the prediction and observation of the order parameter. Very often, the superconducting boundary in $(T, P)$ phase diagram is mainly determined by resistivity measurements. However recent specific heat \cite{Knebel2004} and NMR experiments under pressure \cite{Kitaoka2005} show clearly that the coexistence of superconductivity and antiferromagnetism is not a general rule whatever is the values of $T_N$ or $T_C$. As it was done in the past \cite{Blount1990} and recently \cite{Kivelson1998}, the simplest approach is to consider two coupled order parameters for AF and S via a coupling term of strength $g$ for the Landau free energy. If  $g \geq 0$, superconductivity and AF compete. As indicated in figure (figure \ref{figure4}), with the intensive $P$ variable, a bicritical point exists with a first order transition from AF to S phases. If $g \leq 0$, a tetracritical point will occurs (figure \ref{figure5}) and each phase enhances the other. A so called SO5 theory has the aim to unify theses two basic states by a symmetry principle \cite{Demler2004}. Of course, the constant $g$ may depend on pressure. Thus a Ginzburg Landau approach is a first step to underline basic possibilities.

In Ce heavy fermion compounds at least for three dimensional  Fermi surface, $T^{max}_{C}$ is one order of magnitude smaller than the corresponding maxima  $T^{max}_{N}$ observed for AF. The coexistence of S and AF has been observed below $P_C$ but at the opposite no example has been reported with $T_N \leq T_C$ in a sharp contrast with the conventional case of REMo$_6$S$_8$. The intuitive idea is that if superconductivity appears first on cooling, a large gap will be open in the main part of the FS and the drop of the density of states at the Fermi level is disadvantaging for AF. In HFC, as the characteristic energies are low and the corresponding magnetic field ($H_C$, $H_K$, $H_{C2}$) in the 10 T range, it may be possible to move the domain of the stability of one phase for $g \geq 0$.

To illustrate the physics of HFC superconductors we have selected two examples of cerium AF HFC which becomes superconductors at $P \approx P_C$ : CeIn$_3$ and CeRhIn$_5$ and one example of ferromagnetic superconductors UGe$_2$. Before their presentation let us stress the key role play in the discovery of new materials.

\section{Materials}

Examples of superconductor HFC of 4f electrons exist mainly for cerium compounds where also large FS with 4f itinerant electrons have been observed. The simplicity of Ce HFC is that in the 4f$^1$ configuration of the trivalent state, it is a Kramer's ion with a total angular momentum J = 5/2. For $P \leq P_V$, the crystal field leads to a doublet ground state. The formation of a single ground state can only be due to a Kondo mechanism or more generally to the Fermi statistics. After the unexpected discovery of superconductivity in CeCu$_2$Si$_2$  at ambient pressure with $T_C \approx  0.6$ K \cite{Steglich1979}, the main steps were : 
\begin{itemize}
\item the unusual $P$ dependence of $T_C$ \cite{Bellarbi1984,Thomas1996} in CeCu$_2$Si$_2$ followed by the clear evidence in CeCu$_2$Ge$_2$ that the superconductivity appearance is coupled to $P_C$ \cite{Jaccard1992},
\item the discovery  of the superconducting domain in CePd$_2$Si$_2$ and CeIn$_3$ centered on $P_C$ \cite{Mathur1998}, 
\item the high $T_C$ reached in 115 Ce compounds which have a low dimensional character \cite{Hegger2000} \cite{Thompson2001}. That enhances $T_C$ and its maxima T$^{max}_{C}$ becomes comparable to T$^{max}_{N}$. 
\item Recently a supplementary perspective was given by the appearance of S in the non centrosymmetric crystal of CePt$_3$Si \cite{Bauer2004}.
\end{itemize}

Despite Yb HFC are often described as the hole analog of the Ce cases, no superconductivity has yet been reported. For Pr HFC where the valence fluctuation will occur between the 4f$^2$ and 4f$^1$ configuration, superconductivity has been detected in the PrOs$_4$Sb$_{12}$  skutterudite \cite{Bauer2002} however the pairing may be due to quadrupolar excitons created by the large dispersion of the crystal field excitation between singlet and excited crystal field levels \cite{Kuwahara2005}.
In 5f HFC (mainly U or Pu), the 5f electrons can overlap. So the coexistence of magnetism (AF or F) with unconventional superconductivity can occur even for $P \leq P_C$ since large magnetic fluctuations may exist already. Highly documented cases are UPd$_2$Al$_3$ for AF \cite{Thalmeier2004} and UGe$_2$ for FM \cite{Flouquet2005}. UBe$_{13}$ \cite{Ott1983} looks as a dense Kondo lattice rather similar to CeCoIn$_5$ described later. URu$_2$Si$_2$ ($T_N \approx 17$ K and $T_C = 6$ K) and UPt$_3$ ($T_N = 6$ K and $T_C = 0.6$ K) exhibit  the coexistence of exotic magnetism with tiny ordered magnetic moments  and superconductivity  \cite{Flouquet2005}. The double superconducting transition of UPt$_3$ at ambient pressure and $H = 0$ was one of the first macroscopic evidence that the order parameter cannot be a simple scalar in SCES  \cite{Fisher1989}.
In many cases, excellent crystals of HFC were grown with (i) large electronic mean free path $\ell$ for the observation of quantum oscillations and for the achievement of the required clean limit ($\ell \geq \xi_0$) for the existence of unconventional superconductivity,  (ii) large size to allow complementary experiments such as inelastic neutron scattering, NMR or sensitive macroscopic measurements as thermal expansion or magnetostriction, (iii) nice shiny surface to realise good thermal contact and thus excellent thermalisation to the low temperature bath and precise thermal transport data. The HFC physics is complex with the interplay of different couplings but clear situations are now well identified.

\section{Antiferromagnetism and superconductivity : CeIn$_3$ and CeRhIn$_5$}

At $P = 0$, the cubic lattice CeIn$_3$ is a AF HFC with $T_N$ = 10 K, a sublattice magnetisation $M_0 = 0.5 \mu_B$ at $T \rightarrow 0$ K and a $k_0 = 1/2,1/2,1/2$ propagation vector \cite{Laurence1980,Benoit1980}. The crystal field ground state is the doublet $\Gamma_7$. The spin dynamic observed by neutron scattering shows a $T_K \approx 10$ K at low temperature, a crystal field splitting $C_{CF} \approx 10$ meV and below $T_N$ a quasi-elastic line and damped spin wave \cite{Knafo2003}. Neutron diffraction experiments under pressure suggest that AF may collapse for $P_C = 2.6$ GPa \cite{Morin1988}. The analysis of the resistivity shows that at $P_C$, $k_B T_K \approx C_{CF}$ i.e $P_C$ and $P_V$ coincide \cite{Knebel2002}.

\begin{figure}[b]
\centering
\includegraphics[scale=0.5,clip]{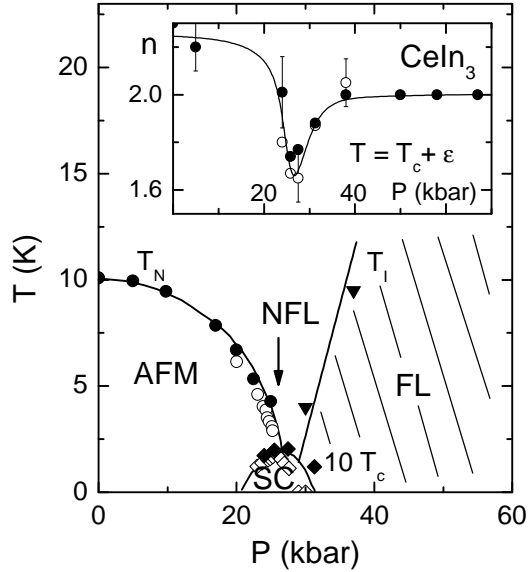}
\caption{Phase diagram of CeIn$_3$. $T_N$  indicates the N$\acute{e}$el temperature, $T_I$ the crossover temperature to the Fermi liquid regime. The superconducting transition temperature $T_C$  is scaled by a factor 10. The exponent $n$ of the pressure dependence of the resistivity ($\rho \approx  T^n$) is shown in the insert. The minimum of the exponent $n$ in the temperature dependence $T^n$ of the resistivity occurs close to the critical pressure $P_C$ (\cite{Knebel2002}) The Fermi liquid in the PM state corresponds to $n = 2$.}
\label{figure6}
\end{figure}

\begin{figure}[t]
\centering
\includegraphics[scale=0.6,clip]{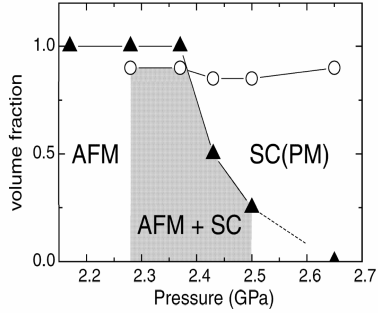}
\caption{NQR data of Kawasaki et al.  on CeIn$_3$ \cite{Kawasaki2004}: volume fraction of AF and S states as a function of $P$.
}
\label{figure7}
\end{figure}

Experiments on a high quality crystal in Cambridge (residual resistivity $\rho_0 \approx 1 \mu\Omega$cm) \cite{Mathur1998} shows that the superconductivity occurs in a narrow $P$ range around $P_C$ with $T^{max}_{C} = 200$ mK (figure \ref{figure6}). Confirmation were found in Osaka \cite{Kobayashi2001} and Grenoble \cite{Knebel2002}. The large initial slope of $H_{C2}(T)$ at $T_C$ proves that the heavy particles themselves condensate in Cooper pairs in agreement with the first observation made for CeCu$_2$Si$_2$ ($T_C \approx 0.6$ K) at $P = 0$ two decades ago \cite{Steglich1979}. Nuclear quadrupolar resonance (NQR) on the In site were very successful to study the spin dynamics notably in the AF and S assumed coexisting regime \cite{Kohori2000}, \cite{Kawasaki2001}. Recent experiments indicate \cite{Kawasaki2004} that the second order nature of the QCP must be questioned as two NQR signals (AF and PM) appears just below $P_C$ (figure \ref{figure7}).  Evidence for the unconventional nature of the superconductivity in both phases is given by the temperature variation of the nuclear relaxation time $T_1$ which follows the $1/T_1 \approx T^3$ law reported for many unconventional exotic superconductors with line of zeros.

The link between S and AF was recently boosted with the Los Alamos discovery of superconductivity in the so called 115 tetragonal cerium compounds like CeRhIn$_5$, CeIrIn$_5$, CeCoIn$_5$ \cite{Thompson2001}. A planar anisotropy is induced by inserting in CeIn$_3$ a single layer of MIn$_5$. The local magnetic anisotropy is weak. The Fermi surface is dominated by slightly warped cylindrical sheets whatever is the localisation of the 4f electrons \cite{Shishido2002}. The gold mine of these compounds are that they cover all the possibilities of interplay between AF and S : - CeRhIn$_5$ which is an AF at $P = 0$ with $T_N = 3.8$ K, becomes PM at $P^{*}_{C} \approx 1.9$ GPa \cite{Fisher2002}. The two others (CeIrIn$_5$, CeCoIn$_5$) are already on the PM side ($P_C \leq  0$) and superconductors at $T_C = 0.4$ K and 2.3 K \cite{Thompson2001}.

As for the other Ce HFC, in CeRhIn$_5$ superconductivity emerges near $P^{*}_{C}$; its maxima $T^{max}_{C} = 2$ K is not too far $T^{max}_{N} = 3.8$ K.  We will focus here on the coexistence of S and AF near $P^{*}_{C}$. Extensive works can be found either under pressure \cite{Fisher2002} or by alloying i.e studies on CeRh$_{1-x}$Co$_x$In$_5$ and CeRh$_{1-x}$Ir$_x$In$_5$ \cite{Zapf2001}, \cite{Pagliuso2001}. 

\begin{figure}[h]
\centering
\includegraphics[scale=0.6,clip]{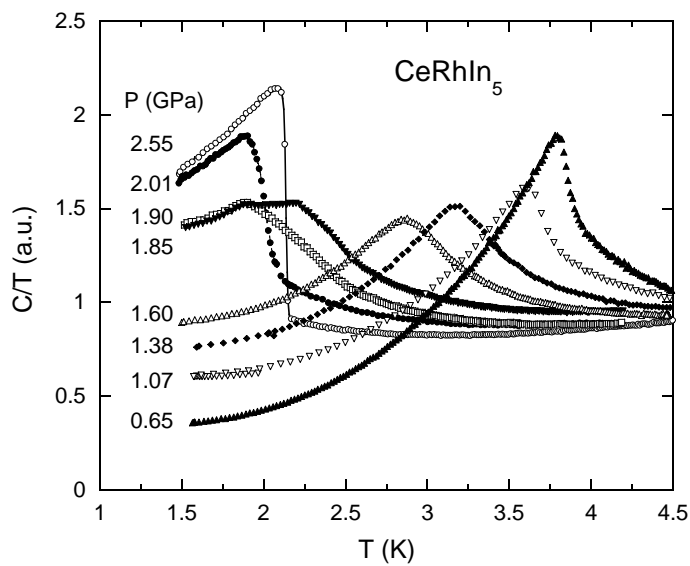}
\caption{Specific heat anomalies of CeRhIn$_5$ at different pressure. At P $\approx$ 1.9 GPa, there is a superposition of tiny superconducting and magnetic anomalies (Knebel et al 2004).}
\label{figure8}
\end{figure}

\begin{figure}[h]
\centering
\includegraphics[scale=1.2,clip]{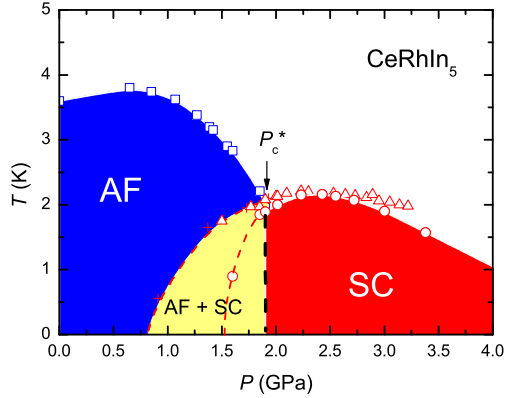}
\caption{$(T, P)$ phase diagram of CeRhIn$_5$. Squares and circlescorrespond to the specific heat anomalies at $T_N$ and $T_C$; triangles to $T_C$ in susceptibility. The AF anomaly disappears suddenly at $P_C \approx 1.9$ GPa. Gapped superconductivity is observed above $P_C$. Inhomogeneous ungapped superconductivity may occur below $P_C$  (white domain) (\cite{Knebel2004}).}
\label{figure9}
\end{figure}

To clarify the situation, careful ac calorimetric measurements to detect $T_N$ and $T_C$ in excellent hydrostatic conditions (Argon pressure transmitting medium) with in situ $P$ tuning at low temperature have been performed \cite{Knebel2004}. Qualitatively, the important feature is that clear AF specific heat anomalies are observed below $P_C^{*} \approx 1.9$ GPa and a clear superconducting anomalies one just above $P_C$ (figure \ref{figure8}). Thus AF and superconductivity seems here to repeal each other. However tiny AF or superconductivity anomalies are detected just below $P_C$. The domain of homogeneous coexistence of AF and gapped superconducting phases may not exist. Indeed, ac susceptibility experiments on a sample coming from the same batch show only a broadened diamagnetism below $P^{*}_{C}$ ; furthermore at higher temperatures than the superconducting specific heat anomaly (figure \ref{figure9}). A sharp diamagnetic transition occurs for $P \geq P_C$ in coincidence with the superconducting specific heat anomaly. It was proposed from NQR measurements that the observation of the inhomogeneity may not be a parasitic effect but an intrinsic property of a new gapless superconducting phase of parity and odd frequency pairing \cite{Kawasaki2003,Fuseya2003}. At least in the NQR studies, simultaneous measurements of ac susceptibility show that its temperature derivative has its maxima at $T_C$ far lower than the previous determination of $T_C$ ($\rho$) by resistivity. No track of AF has been detected above $P_C$ but let us stress that its detection is difficult: in resistivity by the short circuit of the superconducting component and in calorimetry by the collapse of the specific heat magnetic anomaly with the sublattice magnetisation. The observation of an extra inhomogeneous component is yet not resolved above $P^{*}_{C}$. For an intensive variable as the pressure and a single type of particle, only a phase separation is predicted (figure \ref{figure4}). In the complex HFC matter extra effects can occur. Furthermore an important experimental point is that in high pressure diamond anvil as in pressure clamp, the condition to work at constant pressure cannot be certified.

\begin{figure}[h]
\centering
\includegraphics[scale=1.2,clip]{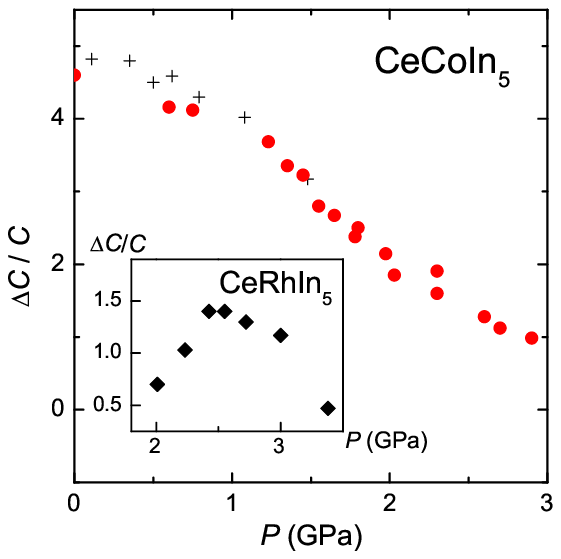}
\caption{Jump of the superconducting specific heat anomaly of CeCoIn$_5$ normalised to the value just above $T_C$ (Knebel et al. \cite{Knebel2004}). The insert shows the variation observed in CeRhIn$_5$.}
\label{figure10}
\end{figure}

We conclude that homogeneous AF and gapped S appear here antagonist. AF collapses through a first order transition at $P^{*}_{C}$ due to the appearance of gapped S. The new phenomena is that under magnetic field so far $H_C \geq H_{C2}(0)$, AF will shift to higher pressure as indicated in figure \ref{figure4}. Basically at $H_{C2}(0, P_C$), $T_N(P_C)$ will reach zero. Extrapolating from other HFC examples (see CeIn$_3$), $T_N (P_C) = 0$ corresponds roughly to the linear extrapolation of $T_N$ i.e here to $P_C = 2.4$ GPa. The amazing features are that it corresponds to the maxima of $T_C$, to the pressure where in de Haas van Alphen experiment a drastic change of the FS occurs with a localisation ($P \leq P_C$) and delocalisation ($P \geq P_C$) of the 4f electrons \cite{Shishido2005} and to the maximum of the jump in the specific heat $\Delta C/C$ at $T_C$ (see insert in figure \ref{figure10}). We suspect that the field change of the ($T, P$) phase diagram AF - S is the origin of the new induced superconducting phase of CeCoIn$_5$.

Figure 10 (figure \ref{figure10}) represents the pressure variation of $T_C$ of CeCoIn$_5$ and of the specific jump $\Delta_C/C(T_C)$ up to 3 GPa. $T_C  (P)$ reaches its maximum for $P = 1.5$ GPa while the specific jump at $T_C$ continuously decreases under $P$. Neglecting strong coupling effects, the jump normalised to the value of the effective mass at $T = 0$ K must be universal i.e related to the strength of $m^* T_C$. Thus the effective mass decreases gradually under pressure. In CeRhIn$_5$ the maximum of $m^*$ seems to occur at $P_c$.

\begin{figure}[h]
\centering
\includegraphics[scale=0.6,angle=90,clip]{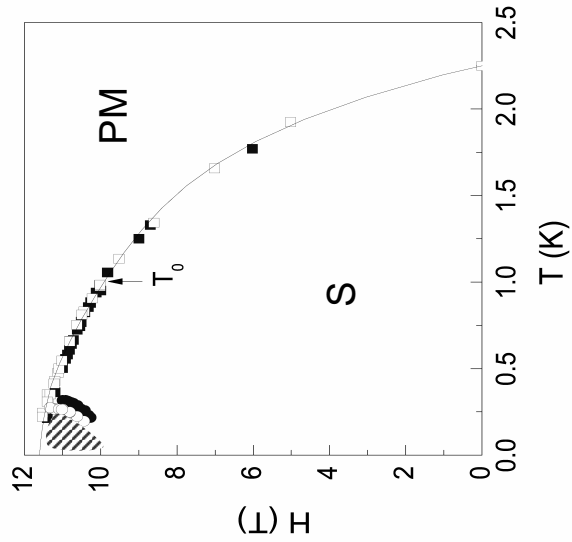}
\caption{$H-T$ phase diagram of CeCoIn$_5$ with both $H \parallel [110]$ (filled symbols) and $H\parallel [100] $(open symbols). (Circles) and (full squares) indicate the assumed $T_{FFLO}$ anomaly for $H \parallel [110]$ and $H \parallel [100]$, respectively. \cite{Bianchi2003a} The hatched domain is the new mixed superconducting phase which may be not a FFLO state. }
\label{figure11}
\end{figure}

In magnetic field  (figure \ref{figure11}) the new features are : (i) a crossover from second order to first order in $H_{C2}(T)$ at $T_0$, (ii) a new high field phase in a restricted low temperature domain \cite{Bianchi2003a,Radovan2003}. Conservative explanations come from predictions made three decades ago : change from 2$^{\rm nd}$ order to 1$^{\rm st}$ order if $H_{C2}(T)$ is dominated by the Pauli limit at low temperature \cite{Saint-James1969} as well as appearance of a new modulated S phase referred as the FFLO state from the work of Fulde Ferrel \cite{Fulde1964} and Larkin Ovchinnichov \cite{Larkin1965} below $T_{FFLO} \approx 0.56 T_C$.

An alternative explanation is that at ambient pressure, CeCoIn$_5$ is located just above $P^{*}_{C}$ but below $P_C$ where $T_C$ may reach its maxima and $T_N$ a vanishing value if the superconductivity will not appear. Applying a magnetic field will lead at $T_0$ to recover the condition $T_C (H) = T_0 = T_N$ $(P = 0, H)$ value of $T_N$ without superconductivity equal basically to the fictitious value at $H = 0$ since $H_C (T = 0) \geq H_{C2}(0)$ so far $T_N$ has not collapsed.

A strong support for this picture is that, crossing through $H_{C2}$, the properties near a magnetic QCP are recovered \cite{Paglione2003,Bianchi2003b}. The Fermi liquid regime is restored continuously to high temperatures with increasing $H$ as observed in CeNi$_2$Ge$_2$ \cite{Gegenwart2003} and in YbRh$_2$Si$_2$ \cite{Tokiwa2004} HFC considered to be almost right at QCP for $P = 0$. 

The new $H$ matter implies a fancy coherence between superconducting and normal component (may be AF or F). Its narrow domain needs further microscopic investigations or analysis. The underlining possibility will be if the magnetic interaction is also modified. As pointed out, for weak local magnetic anisotropy as it happens for CeCoIn$_5$, ferromagnetic and antiferromagnetic fluctuations will compete above $P_C$ i.e near $H_{C2}(0)$. Recent Nernst and dHvA experiments show that the crossover Kondo field $H_K \approx 23$ T occurs far above $H_{C2}(0)$ \cite{Sheikin}. Let us also stress that, as the disorder will change the first order nature of the AF - S boundary, it will add extra effects than those predicted for unconventional superconductivity (see recently \cite{Tanatar}.

\section{UGe$_2$ : a ferromagnetic superconductor}

The appearance of superconductivity in UGe$_2$ \cite{Saxena2000} was a surprise. The relevance of ferromagnetic fluctuations \cite{Anderson}, \cite{Nakajma1973} for anisotropic BCS states was illustrated by the p wave superfluidity of liquid $^3$He \cite{Osheroff1972}. The p wave $T_C$ in nearly ferromagnetic metals was first calculated in 1971 \cite{Layzer1971} and in both FM and PM phases in 1980 \cite{Fay1980}.

At P = 0, UGe$_2$ is a ferromagnet with Curie temperature $T_{\rm Curie} = 54$ K, $M_0$ = 1.48 $\mu_B$/U atom far lower than the full moment near 3$\mu_B$ of the free trivalent or tetravalent uranium configuration \cite{Huxley2001}. Its residual $\gamma$ linear tempeature term of $C$ is 35 mJ mole$^{-1}$K$^{-2}$. Under pressure, T$_{Curie}$ decreases as $M_0$. However as reported for many itinerant ferromagnet, FM disappears via a first order transition with a jump $\Delta  M_0$ = 0.8 $\mu_B$/U \cite{Huxley2001}, \cite{Pfleiderer2002}.

The assertion that even at $P = 0$, UGe$_2$ is itinerant was based on the already described $P$ collapse of $T_{\rm Curie}$ and $M_0$ \cite{Pfleiderer2002}, on band calculations with their success to explain dHvA oscillations \cite{Settai2002}, \cite{Terashima2001}, and on the spin dynamics determined by neutrons scattering with a finite magnetic coherence length $\xi_m$ ($T \rightarrow O$ K) = 24 $\dot{A}$ larger than the typical value found in localised system restricted to atomic distances (6 \AA)\cite{Huxley2003a}.

\begin{figure}[h]
\centering
\includegraphics[scale=0.4,angle=90,clip]{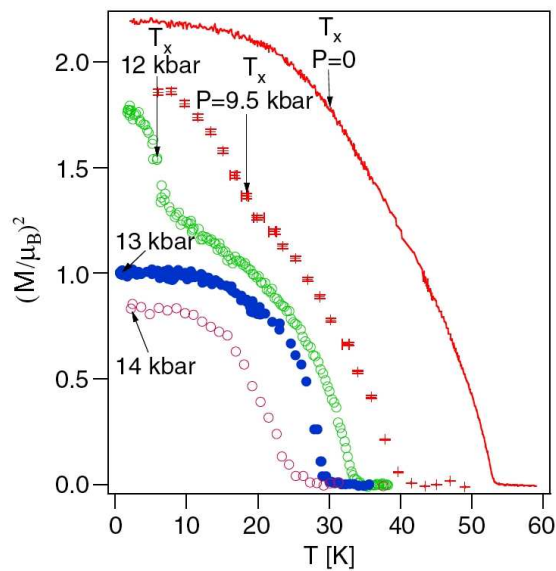}
\caption{The temperature dependence of the ordered moment squared at different pressure deduced from neutron scattering measurements (\cite{Huxley2003b}).}
\label{figure12}
\end{figure}

Figure \ref{figure12} shows the temperature dependence of the ferromagnetic intensity measured by neutron scattering (i.e the square of the ordered moment $M (T)$) at different pressures. Above a critical pressure $P_X = 1.2$ GPa no structure appear in the temperature variation of $M^2 (T)$ while, below $P_X$, at a characteristic temperature $T_X$, a jump appears on cooling. $P$ magnetisation experiments indicate that a discontinuity in $M_0$ appears at $P_X$ at $T \rightarrow 0$ K. Through a first order transition, FM2 transits to FM1 under pressure. The transition FM1 $\rightarrow$ FM2 are easily observed in resistivity measurements when $P$ approaches $P_X$ \cite{Huxley2001}. The competition between two FM phases can be explained in a band theory picture with two phases which differ only by a smooth difference in orbital momentum on the uranium site \cite{Shick2004}. $P$ specific heat measurements analysed via $\gamma T$ contribution plus a $\beta  T^3$ one show a strong increase of $\gamma$ and a maximum of $\beta$ at $P_X$ \cite{Fisher} \cite{Flouquet2005}. Large changes of the FS have been observed between FM2  ($P \leq P_X$) and PM ($P \geq P_C$) , dHvA in FM1  ($P_X \leq P \leq P_C$) are controversial \cite{Settai2002}, \cite{Terashima2001}.

\begin{figure}[h]
\centering
\includegraphics[scale=0.4,clip]{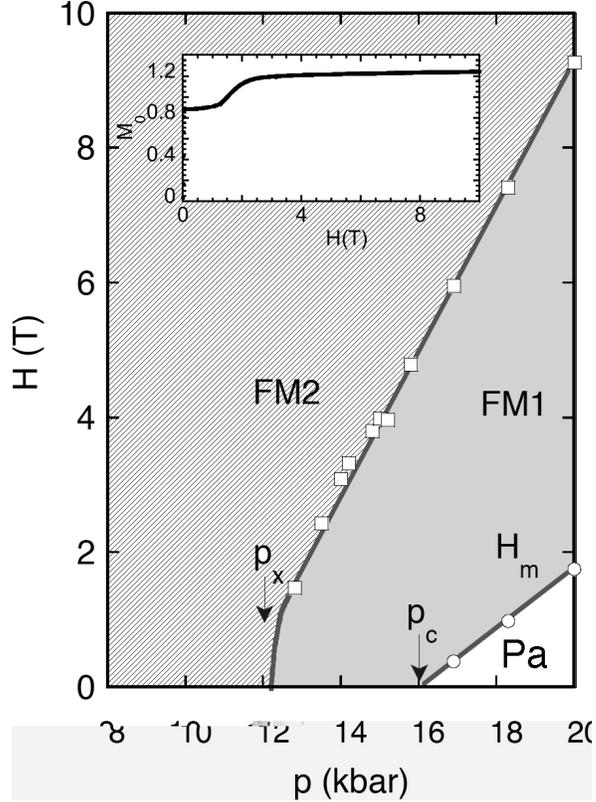}
\caption{$(H, P)$ phase diagram of UGe$_2$ the insert show the jump of $M_0$ in $\mu_B$ at the transition FM$_1$ $\rightarrow$ FM$_2$ at $P_X$ for T = 2.3 K (\cite{Pfleiderer2002}).}
\label{figure13}
\end{figure}

Correlatively to the $P$ induced transition from FM2 to FM1 the magnetic field restores FM2  for $P \geq P_X$ via a metamagnetic transition at $H_X$ and leads also to the cascade PM $\rightarrow$ FM1 at $H_m$ as indicated  (figure \ref{figure13}) \cite{Huxley2001}, \cite{Pfleiderer2002}.

The discovery of superconductivity \cite{Saxena2000} just above 1.0 GPa close to $P_X$ and below $P_C$ inside the ferromagnetic domain is remarkable as superconductivity occurs when $T_{\rm Curie}$ is still high and $M_0$ large : $T^{max}_{C} \approx 700$ mK at $P_X$ when $T_{\rm Curie} = 35$ K and $M_0 = 1.2 \mu_B$ (figure \ref{figure13}). The fact that the superconductivity is not filamentary was first suggested in flux flow experiments and established without ambiguity by the observation of a nearly 30 per cent  specific heat jump at $T^{max}_{C}$ \cite{Tateiwa2001}. Experiments on single crystal achieved in Grenoble \cite{Osheroff1972}, \cite{Huxley2001}, Cambridge \cite{Saxena2000}, Osaka\cite{Tateiwa2001} and Nagoya \cite{Motoyama2001} confirm the S domain at least on materials respecting the condition of a clean limit for superconductivity ($\ell \geq \xi_0$). The ($T, P$) phase diagram of UGe$_2$ is shown (figure \ref{figure13}). Applying a magnetic field in the FM1 - S phase lead to enter in the FM2 - S state for $H = H_X \leq  H_{C2}$ for $P \approx P_X + \epsilon$. The $H_{C2}$ curve can be analysed with equal spin pairing triplet order parameter in good agreement with the large estimation of the exchange field $(M_0 H_{ex} = k_B T_{\rm Curie}) \approx 100$ T while the Pauli limit will be 1 T at $T^{max}_{C} = 0.7$ K.

\begin{figure}[t]
\centering
\includegraphics[scale=0.4,clip]{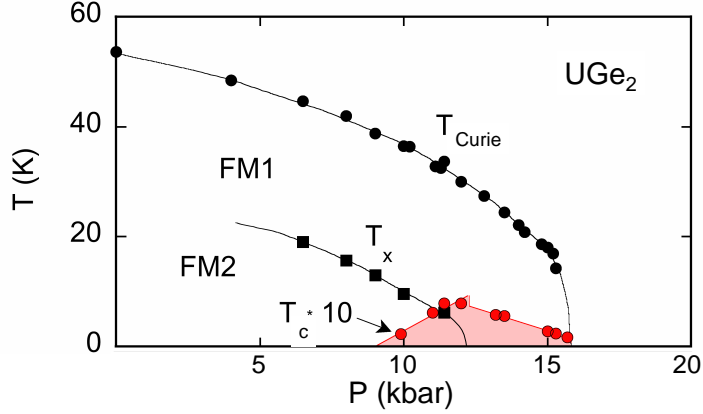}
\caption{$(T, P)$ phase diagram of UGe$_2$. The Curie temperature $T_{\rm Curie}$, the supplementary characteristic temperature $T_X$ which leads to first order transition at $T \rightarrow 0$ K and the superconducting temperature $T_C$ are shown. ($T_C$ scale has been multiplied by 10).
}
\label{figure14}
\end{figure}

\begin{figure}[h]
\centering
\includegraphics[scale=.6]{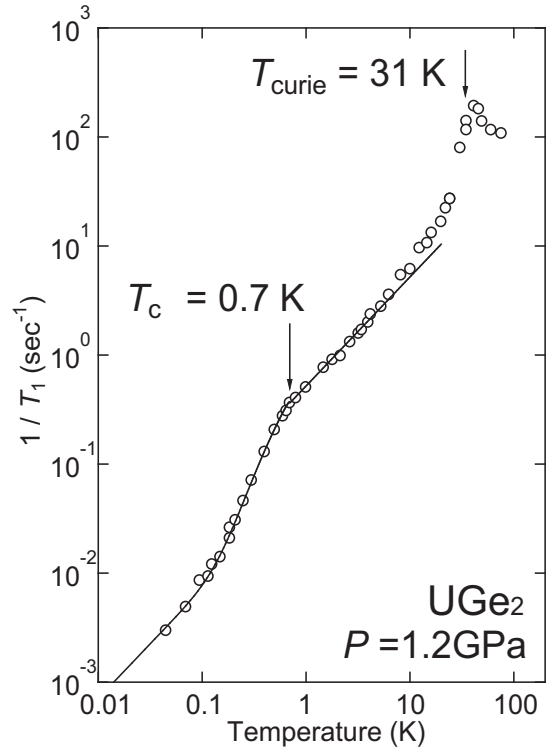}
\caption{At $P=1.2$ GPa, temperature variation of$1/T_1$ with the clear signature of ferromagnetism and superconductivity. The solid line is a calculation assuming an unconventional line node gap \cite{Kotegawa2005}. 
}
\label{figure15}
\end{figure}

Coexistence of superconductivity and ferromagnetism was first verified by neutron scattering \cite{Huxley2001}, \cite{Huxley2003b} and more recently by NQR experiments on Ge sites \cite{Tateiwa2001}. In NQR $1/T_1$ exhibit a peak at $T_{\rm Curie}$ and a change of slope at $T_C$ from a Korringa law to a $T^3$ law characteristic of unconventional superconductors with line of zeros ((figure \ref{figure15}).

Considering the mechanism of superconductivity, obviously the superconducting domain is not centered on $P_C$ but at $P_X$. Specific heat measurements suggest two mechanisms as the $\gamma T$ and $\beta T^3$ terms have quite different $P$ dependences. An interesting proposal is that a charge density wave may occur below $T_X$ \cite{Watanabe2002}. The drop of the resistivity at $T_X$ as well as the coincidence of $T^{max}_{C}$ when $T_X$ collapses is reminiscent of the paramagnet $\alpha$ uranium where $T_X$ is identified as the charge density wave temperature $T_{CDW}$ \cite{Grubel1991,Lander1994}. Furthermore its $T_C$ is also maxima when $T_{CDW} \rightarrow 0$K \cite{Smith2000}. In UGe$_2$ as in $\alpha$ uranium, the U atoms are arranged as zigzag chains of nearest neighbours that run along the crystallographic a axis which is the easy magnetisation. The chains are stacked to from corrugated sheets as in $\alpha$ uranium but with Ge atoms inserted along the b axis. Up to now, no superstructure has been detected below $P_X$. Now that superconductivity is well established, the superconducting states on both side of $P_X$ deserves special attention notably to precise the nature of unpaired electrons (residual $\gamma  T$ in the specific heat).

To realise precise measurements as done on the unconventional superconductor UPt$_3$, the discovery of a FM superconductors at $P = 0$ is an important step. The condition is realised in the orthorhombic URhGe : $T_C = 300$ mK for $T_{\rm Curie} = 10$ K, $\gamma = 160$ mJ mole$^{-1}$ K$^{-2}$ \cite{Aoki2001}. Here again the $T_C$ strength appears not correlated with $P_C$ as, under $P$, $T_C$ collapses already near 4 GPa while $T_{\rm Curie}$ continues to increase linearly with $P$ at least up to 12 GPa (where $T_{Curie} = 18$ K) \cite{Hardy}. The recent success in the growth of single crystals allows to measure $H_{C2} (T)$ along the three main axis. $H_{C2}(0)$ exceeds the Pauli limitation for fields applied along all three crystal axis. Its temperature variation cannot be reconciled with opposite spin pairing but is well described by a single component odd parity polar parameter with a maximum gap parallel to the a axis \cite{Hardy}. 

Despite different attempts no superconductivity has been reported for ferromagnetic Ce HFC close to $P_C$ \cite{Flouquet2005}. It was claimed that ZrZn$_2$ \cite{Pfleiderer2001} will be an ideal ferromagnetic superconductor with $T_C = 0.2$ K. The reproducibility of the effect as its link with the sample quality was rapidly controversial \cite{Flouquet2005}. Recently, it was accepted even by the majority of the discoverers that ZrZn$_2$ superconductivity is not intrinsic \cite{Yelland2005}. Finally it is amazing that $\epsilon$ Fe the high pressure of iron (P $\geq$ 10 GPa) may be a triplet superconductor in its PM phase \cite{Shimizu2001}, \cite{Jaccard2002}.

\section{Conclusion}

Heavy fermion materials give the opportunity to study deeply S phases in an interplay with AF or F. When $T_C$ $\geq$ $T_N$, S and AF appear antagonist. For $T_C \leq T_N$, evidences were given of the coexistence of either AF and ungapped S (CeRhIn$_5$) or a phase separation between AF and PM states both superconducting (CeIn$_3$). In FM materials, superconductivity was discovered only in FM phase with rather large $T_{\rm Curie}$ but presumably in the vicinity of an incipient first order transition which may induce favourable soft modes for S pairing.

Regarding $(T, P)$ phase diagram, an interesting extension is its magnetic field evolution and complementary studies with a weak disorder. In the vicinity of the first order transition at $P^{*}_{C}$ or $P_x$, the inhomogeneous behaviour must be elucidated. For AF systems, the confirmation that a symmetric situation does not occur on PM side needs to be carefully verified. These studies enter in the conjecture of quantum first order transitions at low temperature which up to now have been neglected. As just underlined in the introduction, the accepted consensus of a second order magnetic QCP in HFC may be often not realised \cite{Flouquet2005}.

The superconductivity of HFC has played also an important role in the understanding of unconventional superconductivity : evidence of point and line nodes \cite{Thalmeier2004}, consequence with the magnetic field Doppler shift of the excitations \cite{Suderow1998}, proof that any impurity must be treated by the unitary limit \cite{Pethick1986}, observation of universal law for thermal conductivity \cite{Lee1993} \cite{Joynt2002}, possible occurrence of new mixed state with the possible recovery of exotic cores.

This article is submitted for the Special Issue of the journal {\it Comptes rendus de l'Academie des Sciences: Problems of the  Coexistence of Magnetism and Superconductivity}, edited by A.~Buzdin.

% \appendix

% \section{}
% \label{}

\end{document}